\documentclass{article}
\usepackage{spconf,amsmath,graphicx}
\usepackage{multirow}
\usepackage{hyperref}
\hypersetup{
	colorlinks=true,
	linkcolor=red,
	filecolor=blue,
	urlcolor=blue,
	citecolor=green,
}
\usepackage{url}

\usepackage{amssymb}
\title{SigVIC: Spatial Importance Guided Variable-Rate Image Compression}
\name{Jiaming Liang$^{1}$ \qquad Meiqin Liu$^{1}$ \qquad Chao Yao$^{2}$ \qquad Chunyu Lin$^{1}$ \qquad Yao Zhao$^{1}$\thanks{\emph{Corresponding author: Meiqin Liu.}}}

\address{\small{$^{1}$ Institute of Information Science, Beijing Jiaotong University, Beijing 100044, China} \\ 
\small{$^{2}$ School of Computer and Communication Engineering, University of Science and Technology Beijing, Beijing 100083, China}}

\begin{document}
% \topmargin=0mm
% \ninept
%
\maketitle
\begin{abstract}
Variable-rate mechanism has improved the flexibility and efficiency of learning-based image compression that trains multiple models for different rate-distortion tradeoffs. One of the most common approaches for variable-rate is to channel-wisely or spatial-uniformly scale the internal features. However, the diversity of spatial importance is instructive for bit allocation of image compression. In this paper, we introduce a Spatial Importance Guided Variable-rate Image Compression (SigVIC), in which a spatial gating unit (SGU) is designed for adaptively learning a spatial importance mask. Then, a spatial scaling network (SSN) takes the spatial importance mask to guide the feature scaling and bit allocation for variable-rate. Moreover, to improve the quality of decoded image, Top-K shallow features are selected to refine the decoded features through a shallow feature fusion module (SFFM). Experiments show that our method outperforms other learning-based methods (whether variable-rate or not) and traditional codecs, with storage saving and high flexibility.

% Then, a spatial scaling network (SSN) jointly utilize the spatial importance mask and $\lambda$ to calculate weights in spatial domain for variable-rate.

% , in which a spatial gating unit (SGU) is designed for adaptively learning a spatial importance mask. Then, a spatial scaling network (SSN) takes the spatial importance mask to guide the feature scaling and bit allocation for variable-rate.

% Then, a spatial scaling network (SSN) is utilized to calculate the weights of the learned features in spatial domain, where the spatial importance mask and $\lambda$ are jointly considered.

% A spatial gating unit (SGU) is designed for adaptively learning a spatial importance mask, which is utilized to guide the feature scaling and bit allocation for variable-rate through the proposed spatial scaling network (SSN).

\end{abstract}

\section{Introduction}
\label{sec:intro}

% lossy image compression & Importance of variable-rate
With the explosive demand for storing and sharing images, image compression gradually becomes a vital research field. It can reduce the required storage space and transmission cost of images within acceptable distortion. Recently, many learning-based methods \cite{balle2016enda,balle2018variational,minnen2018joint,lee2018context,cheng2020learned} have outperformed the traditional codecs (e.g. JPEG\cite{wallace1992jpeg}, JPEG2000\cite{skodras2001jpeg}, BPG\cite{sullivan2012overview}). However, these methods rely on training multiple models with fixed rate-distortion (RD) tradeoffs, which leads to a proportionate increase in storage occupation and inflexibility in actual situation. Hence the variable-rate mechanism, which has been available in traditional codecs, is required for reducing the storage space and improving the flexibility.

% of compression models. 

% development of variable-rate methods
Some methods have explored learning-based variable-rate image compression, which typically scale the features in encoder to obtain coarser or finer quantized features and inverse the scaling in decoder. Yang et al.\cite{yang2020variable} design a modulated autoencoder, which scales the internal features for adapting to different RD tradeoffs. Choi et al.\cite{choi2019variable} propose a similar conditional autoencoder with quantization bin-sizes for continuous bit-rates. However, improper selection of bin-size can lead to degradation of RD performance \cite{song2021variable}. Yin et al.\cite{yin2022universal} only scale the bottleneck features, but this scaling strategy is unstable and difficult to adapt to a wide range of RD tradeoffs when training in an end-to-end manner \cite{yang2020variable}. It is noted that, all these methods only involve the channel-wise relationship, without utilizing the spatial importance of images.

% Yin et al.\cite{yin2022universal} only scale the bottleneck features before quantization, but this scaling strategy is unstable when training in end-to-end manner, and cannot adapt to large range of RD tradeoffs \cite{yang2020variable}.

% Yin et al.\cite{yin2022universal} only scale the bottleneck features before quantization, but this scaling strategy is not stable to adapt to large range of RD tradeoffs when training in end-to-end manner \cite{yang2020variable}.

% spatial and sharpness
When the bit-rate of encoding an image is limited, the spatial importance of the image has guiding significance for bit allocation and rate control. The works \cite{song2021variable,gupta2022user} have attempted to utilize the spatial importance, but additional quality map need to be generated for each source image in dataset as input. Moreover, Song et al.\cite{song2021variable} manually pre-define the uniform quality maps for image compression, the spatially different importance is not fully utilized.

% Moreover, the quality maps used for image compression task in \cite{song2021variable} are manually pre-defined and spatial-wise uniformly distributed, which does not take advantage of spatial difference.

% However, this method of generating additional inputs for each image before entering the models is not convenient in practical scenarios.

% ours adaptively/importance guided/stage
% Moreover, as mentioned in a latest review \cite{mishra2022deep}, the sharpness of reconstructed image is considered important.
In this paper, we propose a Spatial Importance Guided Variable-rate Image Compression method named SigVIC, which can adapt to arbitrary bit-rates without additional inputs. Specifically, a spatial gating unit (SGU) is designed to adaptively generate a spatial importance mask of image features. Then, a spatial scaling network (SSN) is proposed to employ the spatial importance mask to guide the generation of spatial scale factors for variable-rate mechanism. Besides, to improve the quality of reconstructed image, we propose a Top-K shallow feature fusion strategy to refine the decoded features through a shallow feature fusion module (SFFM).

% we propose a Top-K shallow feature fusion strategy, where Top-K shallow features are utilized to refine the decoded features through a shallow feature fusion module (SFFM).
% with rich edge and texture information

% experiments :complexity with qmap, artifact of qmap
We evaluate our method on Kodak and CLIC datasets in terms of MSE and MS-SSIM. Experiments illustrate that our method achieves better performance than other variable-rate methods, traditional codecs and some well-known learning-based compression methods. Specifically, by calculating the BD-Rate of MSE results on Kodak dataset, our method saves 17.84\%, 10.85\% and 2.78\% bit-rate compared with BPG\cite{sullivan2012overview}, Song et al.\cite{song2021variable} and Cheng et al.\cite{cheng2020learned}, respectively.

\begin{figure*}[htb]
\begin{minipage}[b]{1.0\linewidth}
  \centering
  \centerline{\includegraphics[width=14.2cm]{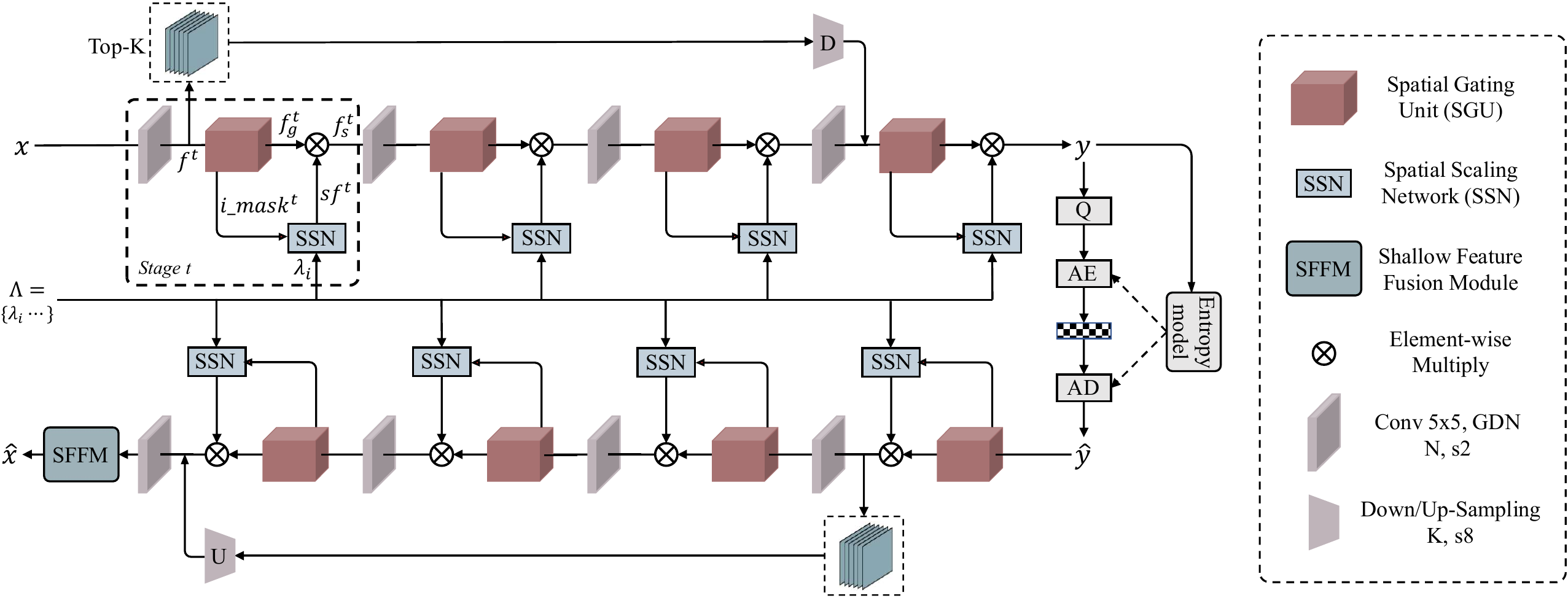}}
  % \centerline{(a) Results 1}\medskip
\end{minipage}
\caption{The overview of our proposed spatial importance guided variable-rate image compression method, SigVIC.}
\label{fig:overview}
\end{figure*}

\section{Proposed Method}
\label{sec:methods}

The overview of our method is depicted in Fig.\ref{fig:overview}, where the encoder and decoder networks consist of several stages. In each stage $t$, a spatial gating unit (SGU) and a spatial scaling network (SSN) are designed to introduce spatial importance mask to guide RD optimization for variable-rate. Besides, a shallow feature fusion module (SFFM) is employed at the end of decoder to refine the decoded features.

% An encoder network transform the input image $x$ to a latent feature $y$, which will be quantized (Q), arithmetically encoded (AE) and decoded (AD) to $\hat{y}$. A entropy model \cite{cheng2020learned} is employed to estimate the bit-rate of $\hat{y}$. Then, a decoder network reconstructs image $\hat{x}$ from $\hat{y}$.

% with a set of rate-distortion (RD) tradeoffs $\lambda_i$, where $\lambda_i$ participates in the scaling of features in each stage $t$, where the features scaling are inverted using the same $\lambda_i$.

% for spatial importance guided variable-rate

\subsection{Spatial Gating Unit}
\label{ssec:method-sgu}

\begin{figure}[htb]
\begin{minipage}[b]{1.0\linewidth}
  \centering
  \centerline{\includegraphics[width=7.0cm]{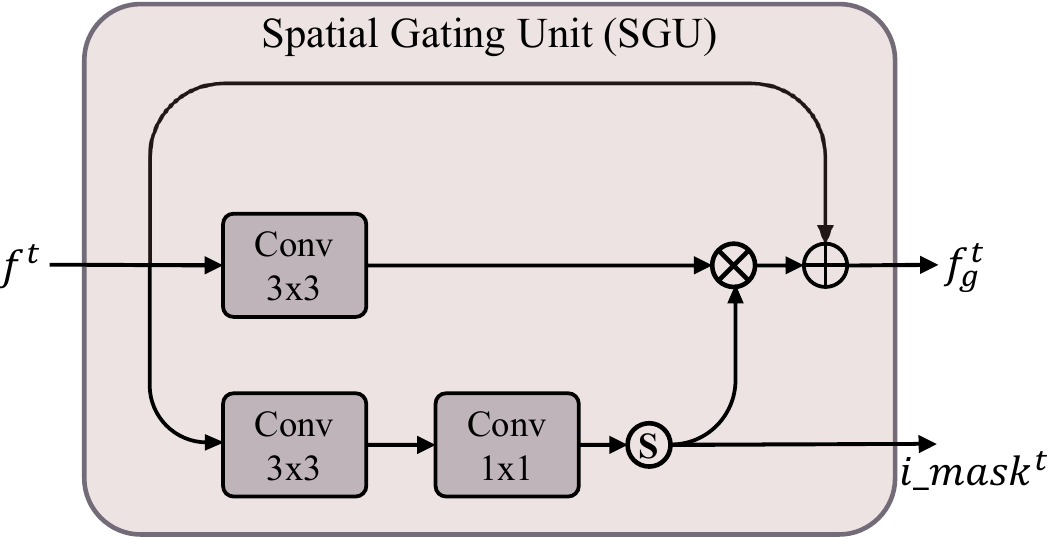}}
\end{minipage}
\caption{The detials of spatial gating unit (SGU).}
\label{fig:sgu}
\end{figure}

For the input features $f^t$ at stage $t$, a spatial gating unit (SGU) is designed for adaptively generating a spatial importance mask $i\_mask^t$, as shown in Fig.\ref{fig:sgu}. To generate $i\_mask^t$, a 3$\times$3 convolution is adopted to initially extract the spatial features from $f^t$. Then, a 1$\times$1 convolution with a sigmoid operation is adopted to produce the weights of $i\_mask^t$. Through the activation of sigmoid, different weights are produced in $i\_mask^t$ for different pixels of $f^t$, and most of them tend to be 0 or 1.

% $i\_mask^t$ produces different weights for different pixels of $f^t$, and most of which will tend to 0 or 1.

\begin{equation}\label{equ:imask}
\setlength\abovedisplayskip{-0.1cm}
\setlength\belowdisplayskip{0.2cm}
\begin{aligned}
&i\_mask^{t} = \text{Sigmoid}(C^{1}(C^{3}(f^{t})))
\end{aligned}
\end{equation}

\noindent By multiplying with equivalent spatial features extracted in another branch using 3$\times$3 convolution, $i\_mask^t$ can play the role of gating unimportant features of $f^t$. After adding with the identity features $f^t$, the gated features $f_g^t$ is obtained as:

\begin{equation}\label{equ:scale}
\setlength\abovedisplayskip{-0.1cm}
\setlength\belowdisplayskip{0.2cm}
\begin{aligned}
&f_{g}^{t} = i\_mask^t \cdot C^{3}(f^t) + f^{t}
\end{aligned}
\end{equation}

\noindent The $i\_mask^t$ is also be passed into our spatial scaling network (SSN) to guide the generation of spatial scale factor.

\subsection{Spatial Scaling Network}
\label{ssec:method-vr}

% To use the spatial importance mask as a guideline for feature scaling and bit-rate control in variable-rate, the  is proposed, as shown in .

\begin{figure}[htb]
\begin{minipage}[b]{1.0\linewidth}
  \centering
  \centerline{\includegraphics[width=7.0cm]{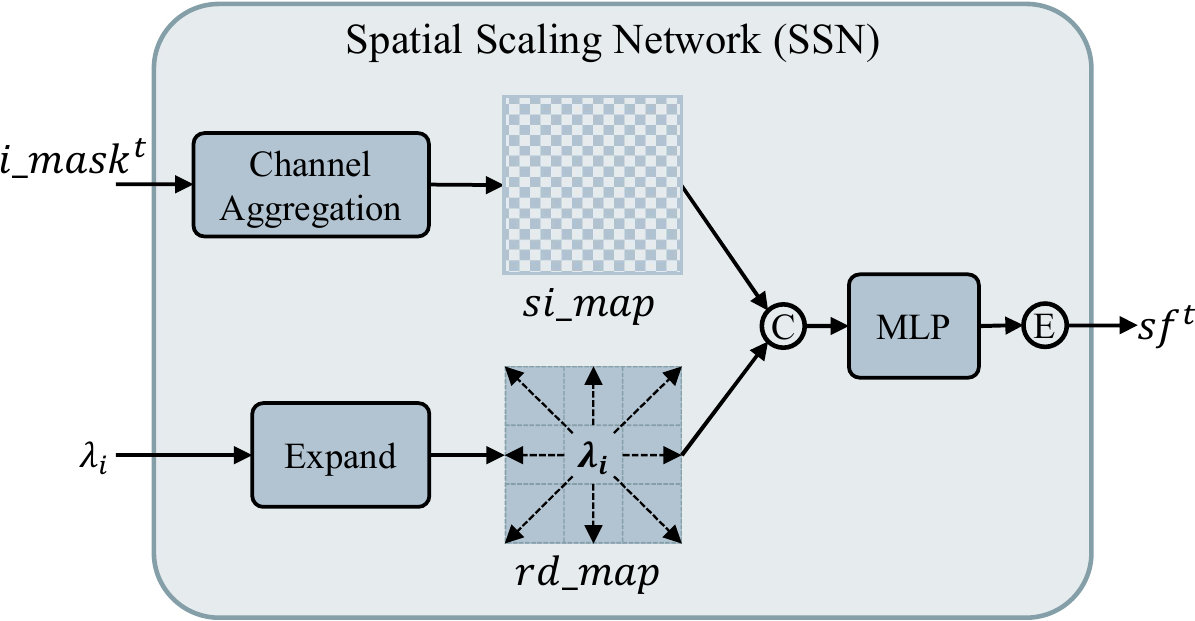}}
\end{minipage}
\caption{The structure of spatial scaling network (SSN).}
\label{fig:ssn}
\end{figure}

\renewcommand{\thefigure}{5}
\begin{figure*}[htb]
\begin{minipage}[b]{0.245\linewidth}
  \centering
  % \centerline{Kodak[MSE]}\medskip
  \centerline{
    \includegraphics[width=4.35cm]{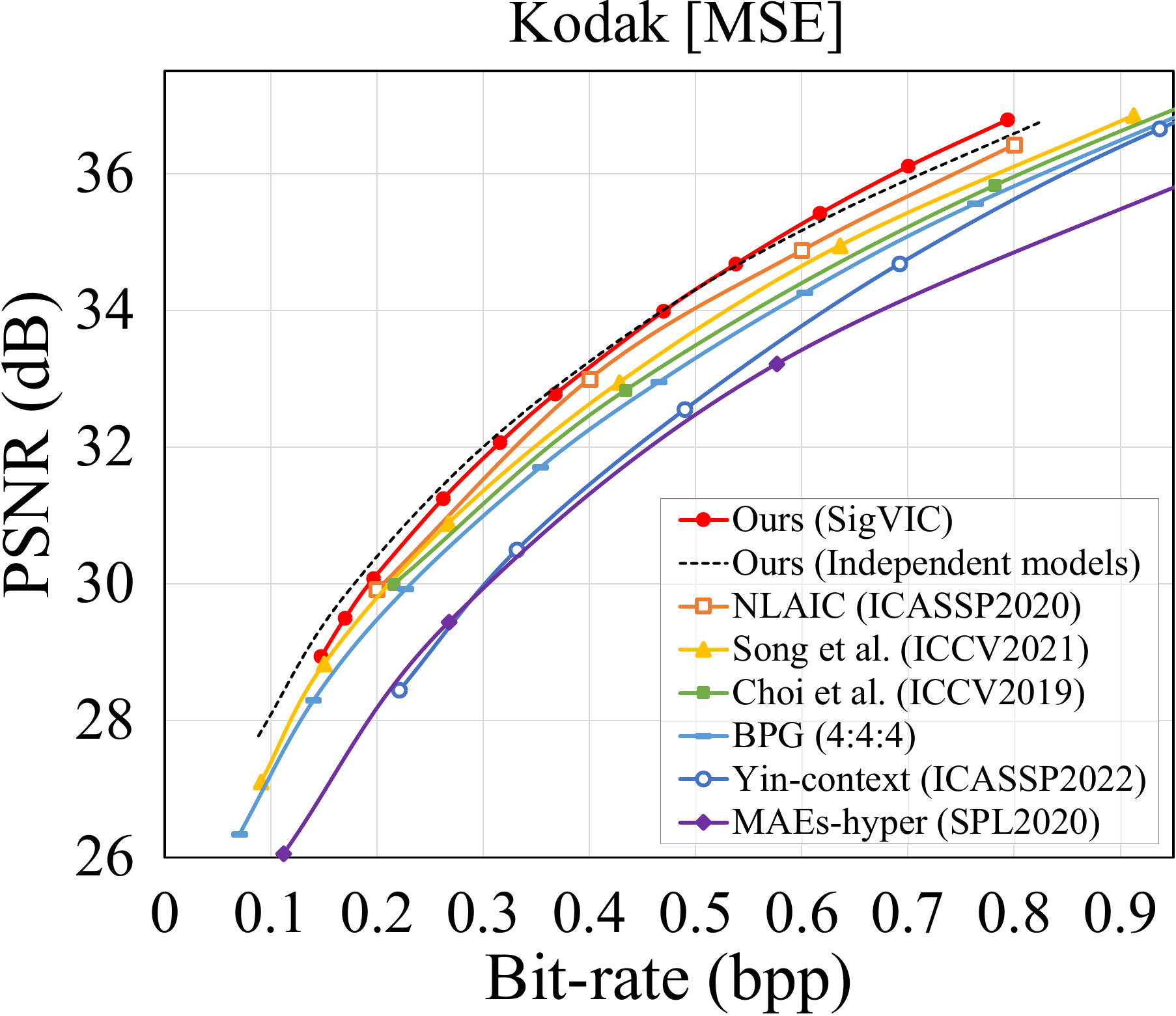}
    % \hspace{-1.5cm}
  }
\end{minipage}
\hfill
\begin{minipage}[b]{0.245\linewidth}
  \centering
  % \centerline{Kodak[MS-SSIM]}\medskip
  \centerline{
    \includegraphics[width=4.35cm]{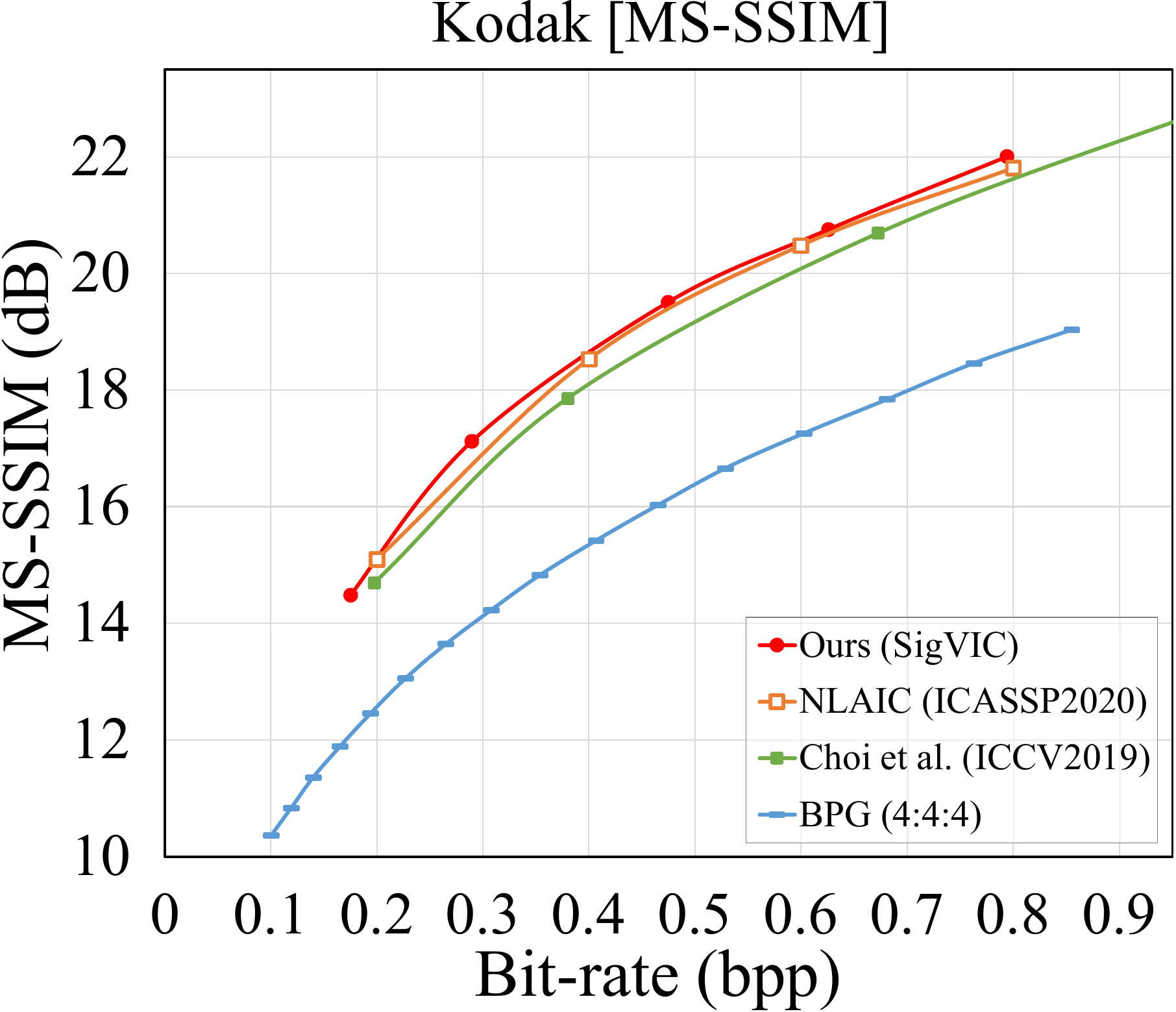}
    % \hspace{-1.0cm}
  }
\end{minipage}
\hfill
\begin{minipage}[b]{0.245\linewidth}
  \centering
  % \centerline{CLIC[MSE]}\medskip
  \centerline{
    \includegraphics[width=4.35cm]{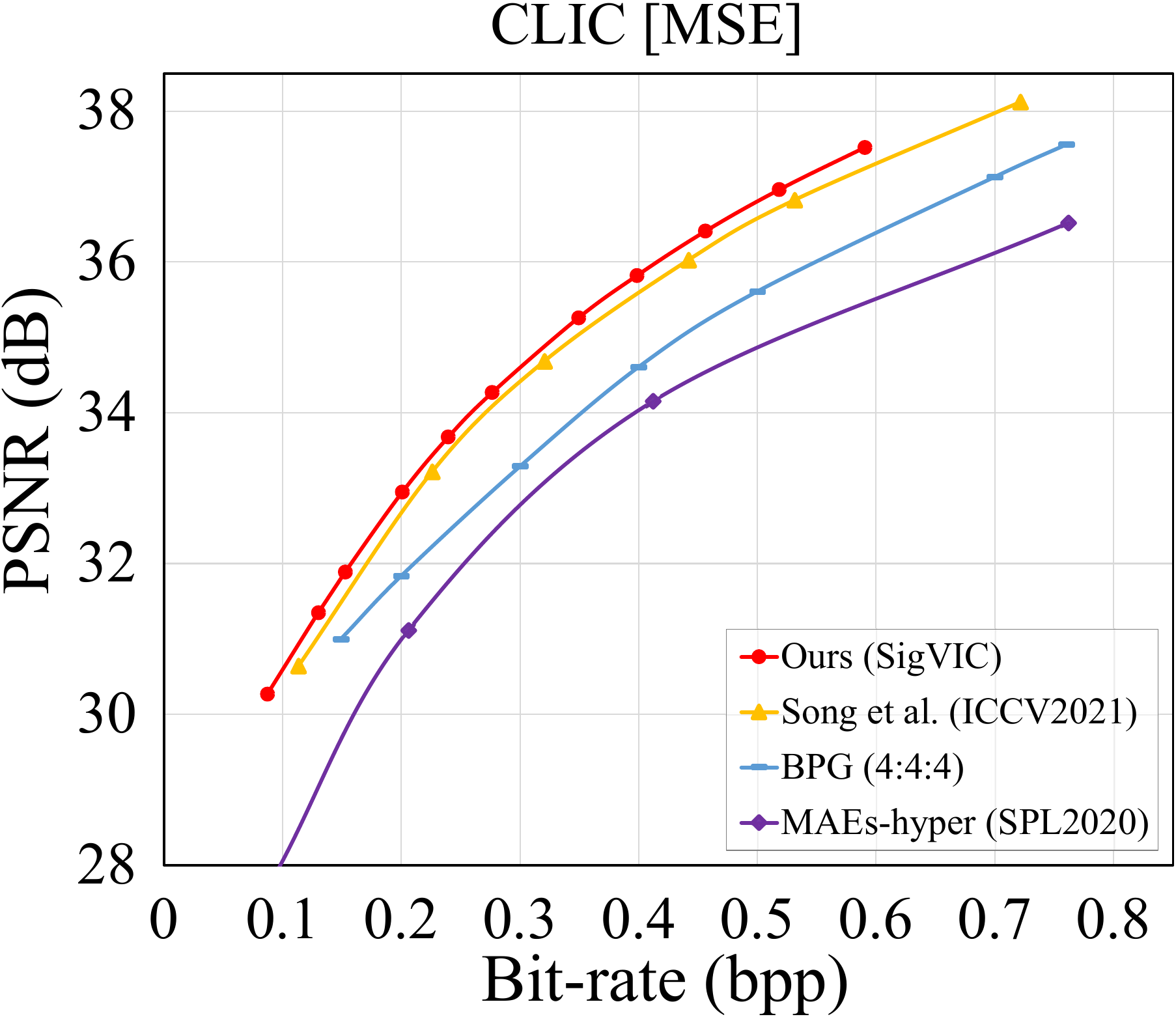}
    % \hspace{-0.5cm}
  }
\end{minipage}
\hfill
\begin{minipage}[b]{0.245\linewidth}
  \centering
  % \centerline{CLIC[MS-SSIM]}\medskip
  \centerline{
    \includegraphics[width=4.35cm]{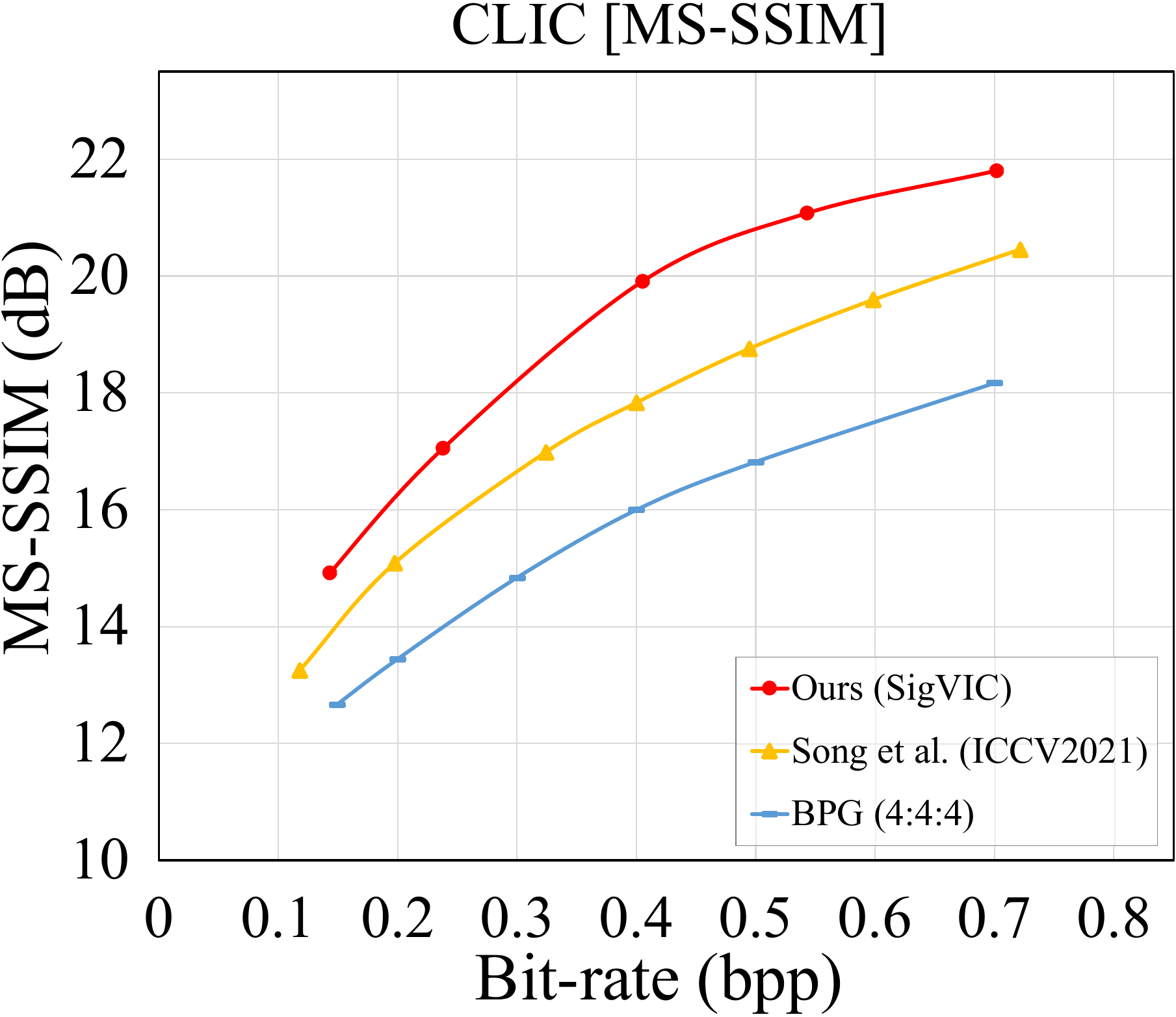}
    % \hspace{0cm}
  }
\end{minipage}
\caption{The RD curves, using MSE and MS-SSIM as the distortion term of loss function, on Kodak and CLIC datasets.}
\label{fig:rd}
\end{figure*}

% SSN has two inputs for generating the spatial scale factor $sf^t$,

% Unlike common variable-rate methods that takes the RD tradeoff $\lambda_i$ as the only. 

% Inspired by MetaSR \cite{hu2019meta} who realizes magnification-arbitrary super-resolution by concatenating coordinate with scale factor to predict weights for different patches. 

% Firstly, both inputs should be converted to the same spatial resolution as the features to be scaled.

The proposed spatial scaling network (SSN) is depicted in Fig.\ref{fig:ssn}. To introduce the spatial importance mask for guiding feature scaling and RD optimization, SSN takes the adaptively learned $i\_mask^t$ and RD tradeoff $\lambda_i$ as the joint inputs. Firstly, the spatial resolution of both inputs should be converted to the same as the gated features $f_g^t$. We aggregate the channel-wise information of $i\_mask^t\in\mathbb{R}^{H \times W \times N}$ and produce a spatial importance map $si\_map\in\mathbb{R}^{H \times W \times 1}$. The constant $\lambda_i$ cannot be directly converted by learning, because $H$ and $W$ are uncertain with different size of input image. Our solution is to tile and expand $\lambda_i$ into a $rd\_map \in \mathbb{R}^{H \times W \times 1}$. After that, we channel-wisely concatenate and input them into a MLP network to generate the spatial scale factor $sf^t$, which has been related to both spatial importance and bit-rate. The MLP consists of two full-connected layers with 64 hidden units and a ReLU in the middle. An exponential function is applied behind for positive outputs, which is beneficial for training process \cite{yang2020variable}.

% The RD curves of Kodak[MSE], Kodak[SSIM], CLIC[MSE], CLIC[SSIM], the distortion of Loss is marked in the square brackets of the title.

% The RD curves of Kodak[MSE], Kodak[SSIM], CLIC[MSE], CLIC[SSIM], what is in the square brackets in title is the distortion of Loss.

\begin{equation}\label{equ:sf}
\setlength\abovedisplayskip{-0.1cm}
\setlength\belowdisplayskip{0.2cm}
\begin{aligned}
&sf^{t} = \text{exp}(MLP(Concat(si\_map,  rd\_map)))
\end{aligned}
\end{equation}

% different positions of $f_g^t$ are scaled to different fineness, and $f_s^t$ is obtained.

% $f_g^t$ is spatially differently scaled to $f_s^t$ with different fineness.

% different positions of feature are scaled to different fineness.

By spatial-wisely multiplying with $sf^t$, different positions of $f_g^t$ are scaled to different fineness and the scaled feature $f_s^t$ is obtained. In this way, the bit allocation for different bit-rates is related to spatial importance after quantization.

% \begin{equation}\label{equ:scaling}
% \begin{aligned}
% &f_s^t = f_g^t \cdot sf^t
% \end{aligned}
% \end{equation}

% % necessary?
% In addition, we do not scale the features in entropy model, because the bit-rate produced here is much smaller and will not affect the performance of the entire network.

\subsection{Top-K Shallow Feature Fusion Module}
\label{ssec:method-ssf}

\renewcommand{\thefigure}{4}
\begin{figure}[htb]
\begin{minipage}[b]{1.0\linewidth}
  \centering
  \centerline{\includegraphics[width=7cm]{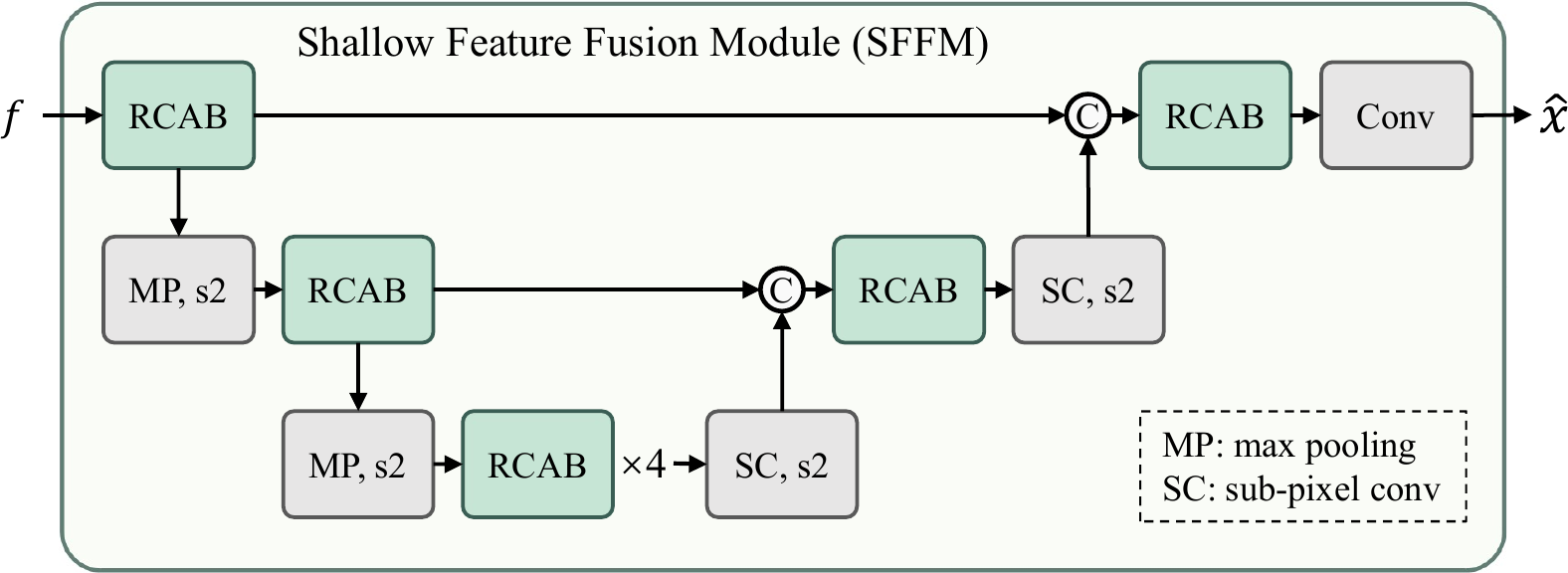}}
\end{minipage}
\caption{Structure of shallow feature fusion module (SFFM).}
\label{fig:sffm}
\end{figure}

To enrich the details of decoded image $\hat{x}$, Top-K shallow features that rich in edge and texture information are selected and incorporated into the encoded feature $y$. These features are restored in decoder to refine the decoded features through a shallow feature fusion module (SFFM). Because most of the information in features is only contributed by a few channels\cite{he2022elic}, we select the most informative $K$ feature maps for balancing the performance and computation.

% necessary ? no
% Since the shallow features are always coarse, which is unfavorable to reducing bit-rate, the last SGU is used for fusing and gating the features. At the decoder side, the Top-K features are restored and up-sampled in a symmetrical way. 

% Unlike common methods directly reconstructing the image, we adopt the SFFM to further integrate the shallow features and decoding features to enrich the details of reconstructed image.

% SFFM is a U-shaped network as shown in Fig.\ref{fig:sffm}. 

% As shown in Fig.\ref{fig:sffm}, SFFM leverages a U-shape network to refine the decoded features. 

SFFM leverages a U-shaped network as shown in Fig.\ref{fig:sffm}. Multiple residual channel attention blocks (RCAB)\cite{zhang2018image} are utilized to provide the learning of channel-wise information. In addition, SFFM can be trained end-to-end in the whole network with only a few parameters.
% 32,64,128   because the numbers of filters are small

% leverages a similar shape with U-Net\cite{ronneberger2015u} to refine the decoded features with shallow features

\subsection{Loss function}
\label{ssec:loss}

% which is adaptively guided by the learned spatial importance mask.
% while the guidance of spatial importance has been completed adaptively.

Our network receives variable $\lambda_i$ for adapting to arbitrary bit-rates, while the guidance of spatial importance has been adaptively completed. The loss function is formulated as:

\begin{equation}\label{eq:loss_new}
\setlength\abovedisplayskip{-0.1cm}
\setlength\belowdisplayskip{0.2cm}
\begin{aligned}
L &= R(\hat{y},\hat{z},\lambda_i) + \lambda_i D(x, \hat{x}, \lambda_i) ~~ (\lambda_i \in \Lambda)
\end{aligned}
\end{equation}

\noindent where $\Lambda$ represents a range between two selected boundary values, containing all the possible values of $\lambda_i$.

% where $\Lambda$ represents a range determined by several boundary values and contains arbitrary values of $\lambda_i$.

\section{Experiments}
\label{sec:exps}

A subset of ImageNet\cite{deng2009imagenet} with 13600 images is used for training our models, which are randomly cropped to patches with resolution of 256$\times$256. The batch size is set to 8, and the models are trained for 2$M$ iterations with Adam optimizer \cite{kingma2014adam}. The learning rate is initially set to 1$\times$10$^{-4}$, and decreased to 1$\times$10$^{-5}$ for the last 0.1$M$ iterations. We train our models using MSE and MS-SSIM loss, where the selected boundaries of $\Lambda$ are respectively (0.0016, 0.045) and (5, 120) for covering wide range of bit-rates. The number of selected shallow features $K$ is empirically set to 32, and the number of filters $N$ of network is set to 192.

% for balancing the computation and performance

\subsection{Rate-distortion (RD) performance}
\label{ssec:rd}

\renewcommand{\thefigure}{7}
\begin{figure*}[htb]
\begin{minipage}[b]{1.0\linewidth}
  \centering
  \centerline{\includegraphics[width=14.2cm]{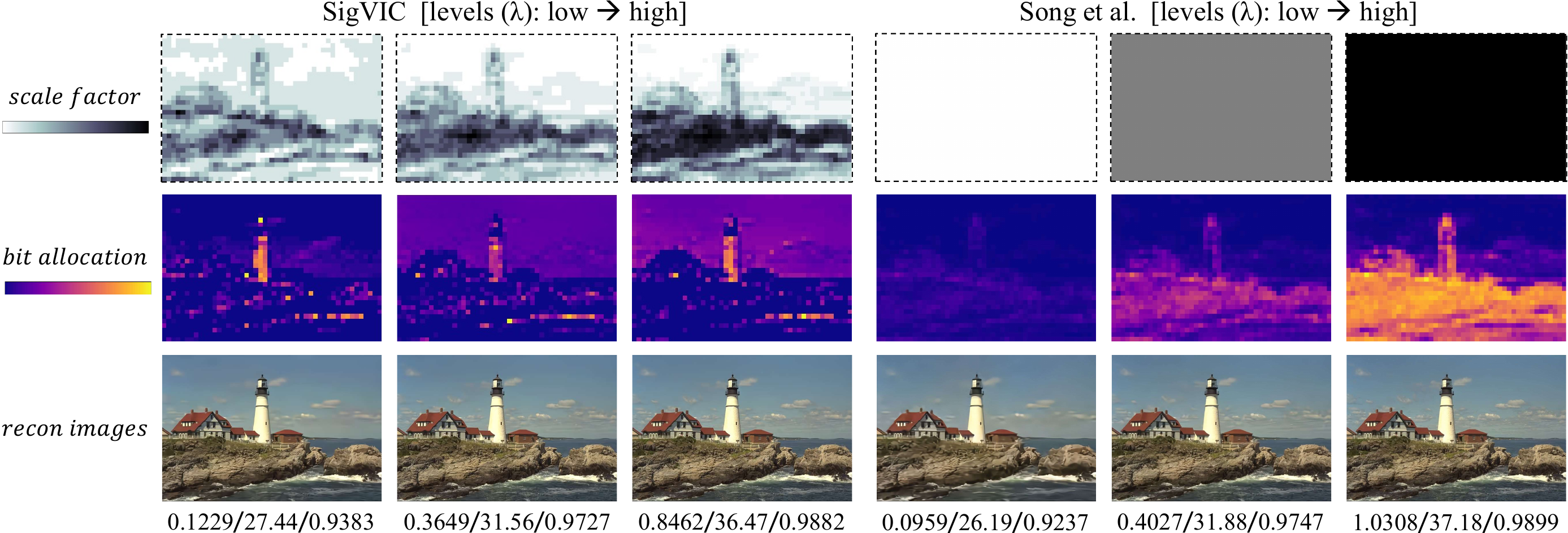}}
\end{minipage}
\caption{The visualizations (including scale factor and bit allocation) of our SigVIC and Song et al.}
\label{fig:vis}
\end{figure*}

We compare the RD curves of our method with other variable-rate compression methods including learning-based methods \cite{yang2020variable,choi2019variable,song2021variable,yin2022universal,chen2020variable} and traditional codec BPG\cite{sullivan2012overview}. As shown in Fig.\ref{fig:rd}, our method outperforms all the other methods in terms of both PSNR and MS-SSIM (values are converted to decibels by $-10log_{10}(1-\text{MS-SSIM})$ for clarity) on Kodak\cite{Kodak24} and CLIC\cite{CLIC} datasets.

\begin{table}[!t]
\vspace{-0.2cm}
\caption{BD-results against BPG on Kodak and CLIC}
\label{table:bd_all}
\centering
\scalebox{0.8}{
    \begin{tabular}{c|cc|cc}
    \hline
    \multirow{2}{*}{Methods} & \multicolumn{2}{c|}{Kodak} & \multicolumn{2}{c}{CLIC}\\
    \cline{2-5}
     & BD-PSNR & BD-Rate & BD-PSNR & BD-Rate\\
    \hline
    MAEs-hyper\cite{yang2020variable} & -0.96 dB & 21.35\% & -0.59 dB & 14.88\%\\
    Yin-context\cite{yin2022universal} & -0.87 dB & 17.30\% & -{}- & -{}-\\
    Choi et al.\cite{choi2019variable} & 0.22 dB & -4.96\% & -{}- & -{}-\\
    Lee et al.\cite{lee2018context} & 0.22 dB & -5.05\% & 0.56 dB & -12.25\%\\
    Song et al.\cite{song2021variable} & 0.37 dB & -8.06\% & 1.05 dB & -21.73\%\\
    NLAIC\cite{chen2020variable} & 0.71 dB & -14.50\% & -{}- & -{}-\\
    Cheng et al.\cite{cheng2020learned} & 0.78 dB & -16.43\% & 1.21 dB & -25.82\%\\
    Ours (SigVIC) & \textbf{0.92 dB} & \textbf{-17.84\%} & \textbf{1.23 dB} & \textbf{-26.16\%}\\
    \hline
    \end{tabular}
}
\vspace{-0.2cm}
\end{table}

% To compare the RD performance more clearly, 

% % 0 row.
% We further compare the BD-results(i.e.BD-PSNR and BD-Rate) with the above methods and some well-known learning-based image compression methods\cite{lee2018context,cheng2020learned} that need to train multiple models for different bit-rates. Higher BD-PSNR and smaller BD-Rate indicate better RD performance, which correspond to more PSNR gains and bit-rate savings. As shown in Table.\ref{table:bd_all}, all the results are calculated against BPG. It can be seen that our method achieves the maximum PSNR gain of 0.92dB and saves the maximum bit-rate of 17.84\% on Kodak dataset, which outperforms all the other methods. On the CLIC dataset with higher resolution, our method obtains higher BD-PSNR of 1.23dB and saves more bit-rate of 26.16\%. Besides, our results are slightly better than Cheng et al.\cite{cheng2020learned}, which achieves comparable performance with VVC\cite{ohm2018versatile}.

% -1 row by modify image(compression)/obviously/gets(obtains)
We further compare the BD-results(i.e.BD-PSNR and BD-Rate) with the above methods and some well-known learning-based compression methods\cite{lee2018context,cheng2020learned} that need to train multiple models for different bit-rates. Higher BD-PSNR and smaller BD-Rate indicate better RD performance, which correspond to more PSNR gains and bit-rate savings. As shown in Table.\ref{table:bd_all}, all the results are calculated against BPG. Obvoiusly, our method achieves the maximum PSNR gain of 0.92dB and saves the maximum bit-rate of 17.84\% on Kodak dataset, which outperforms all the other methods. On the CLIC dataset with higher resolution, our method gets higher BD-PSNR of 1.23dB and saves more bit-rate of 26.16\%. Besides, our results are slightly better than Cheng et al.\cite{cheng2020learned}, which achieves comparable performance with VVC\cite{ohm2018versatile}.

% -1 row by modify bd-results&i.e.

In order to test the robustness of our variable-rate mechanism, we also train 6 independent models without SSN. As shown in the \emph{Kodak[MSE]} results in Fig.\ref{fig:rd}, our SigVIC achieves comparable performance with corresponding independent models. This indicates that our variable-rate mechanism will not affect the performance of compression networks, while improving the efficiency and flexibility.

\subsection{Subjective Performance}
\label{ssec:subj}

Fig.\ref{fig:subj} shows some reconstructed images to evaluate the subjective performance of our method. All of them are at the similar bit-rate of about 0.22 bpp. In the enlarged regions, it can be seen that our method restores more details, such as the words on helmet and pants, the texture of motorcycle tyres and mudguards, and so on.

\renewcommand{\thefigure}{6}
\begin{figure}[htb]
\begin{minipage}[b]{1.0\linewidth}
  \centering
  \centerline{\includegraphics[width=7.6cm]{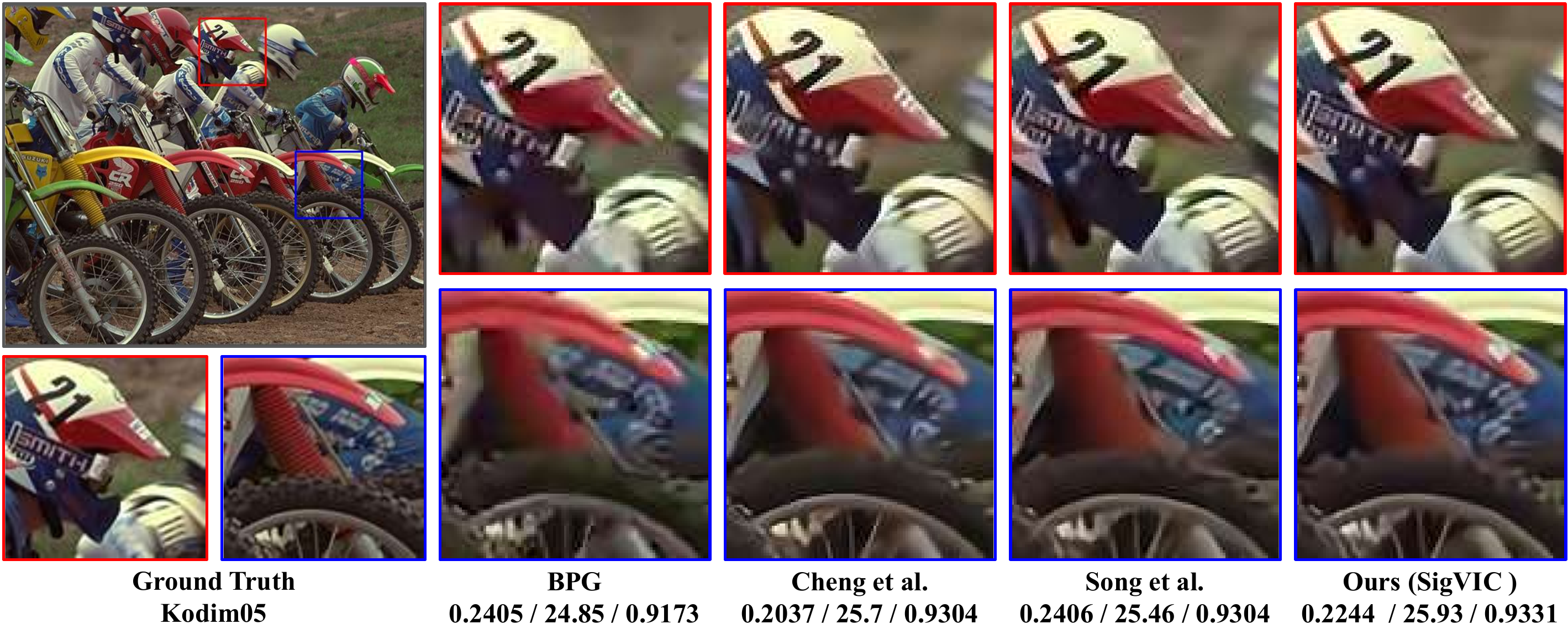}}
\end{minipage}
% \caption{Reconstructed images and details of some methods.}
\caption{Visual quality of reconstructed images.}
\label{fig:subj}
\end{figure}
\vspace{-0.4cm}

\subsection{Visualizations}
\label{ssec:vis}

To evaluate the effectiveness of our spatial importance guided feature scaling and bit allocation, we visualize the scale factors with corresponding bit allocation maps under several levels of RD tradeoff $\lambda$. As shown in Fig.\ref{fig:vis}, our scale factors can adaptively learn the spatial importance of image and scale features of different regions with different weights, while the quality maps of Song et al.\cite{song2021variable} scale different regions using equal weights. Therefore, when the bit-rate is limited, our method can preferentially allocate more bits to informative regions under the guidance of spatial importance.

% In our method, when the bit-rate is limited, more bits can be preferentially allocated to informative regions under the guidance of spatial importance.

\subsection{Ablation Study}
\label{ssec:ablation}

% % 230306(Mon.)
% The results of ablation study are shown in Table.\ref{table:abla}, where the check marks mean corresponding modules are used. The scheme A, in which SSN is used but SGU is not, adopts uniform scale factor for variable-rate mechanism. Compared with Scheme A, Scheme B saves 3.56\% bit-rate, which benefits from the guidance of spatial importance on bit allocation and RD optimization with SGU. Furthermore, scheme C saves 6.92\% bit-rate, which indicates the Top-K shallow feature fusion strategy with SFFM can improve the quality of decompressed image and RD performance.

% where the combination of SSN is T and SGU is F represents using uniform scale factor for variable-rate. 
% When SSN is used but SGU is not, it represents using uniform scale factor for variable-rate mechanism.

We study the effectiveness of SGU and SFFM in our method, the results are shown in Table.\ref{table:abla}. Scheme A adopts a uniform scale factor for variable-rate mechanism, where SSN is used but SGU is not. Compared with Scheme A, Scheme B saves 3.56\% bit-rate, which benefits from the guidance of spatial importance on bit allocation and RD optimization by SGU. In addition, more bit-rate of 6.92\% is saved by Scheme C. This demonstrates that the Top-K shallow feature fusion strategy with SFFM can further improve the quality of decompressed image and RD performance.

% In addition, Scheme C saves 6.92\% bit-rate, which shows that the Top-K shallow feature fusion strategy of SFFM can improve the quality of decompressed images and RD performance.

% Compare with scheme A, scheme B saves 3.56\% bit-rate under the guidance of spatial importance of SGU. 
% Compared with Scheme A, Scheme B saves 3.56\% bit-rate, which is due to the guidance of SGU's spatial importance on bit allocation and RD optimization.

% where the check marks means corresponding modules are used and the cross mark means not.

% and the compression performance gets further improved.

\vspace{-0.1cm}
\begin{table}[!h]
\vspace{-0.2cm}
\caption{Results of ablation study}
\label{table:abla}
\centering
\scalebox{0.9}{
    \begin{tabular}{c|ccc|cc}
    
    \hline
    Schemes & SSN & SGU & SFFM & BD-Rate & Params.\\
    \hline
    A & \checkmark &  &  & 0\% & 15.3M\\
    B & \checkmark & \checkmark &  & -3.56\% & 23.21M\\
    C & \checkmark & \checkmark & \checkmark & -6.92\% & 24.85M\\
    \hline
    
    \end{tabular}
}
\vspace{-0.2cm}
\end{table}
\vspace{-0.175cm}

% \vspace{-0.1cm}
% \begin{table}[!h]
% \vspace{-0.2cm}
% \caption{Results of ablation study}
% \label{table:abla}
% \centering
% \scalebox{0.9}{
%     \begin{tabular}{c|ccc|cc}
    
%     \hline
%     Scheme & SSN & SGU & SFFM & BD-Rate & Params\\
%     \hline
%     A & \usym{2713}\usym{1F5F8} & \usym{2717} & \usym{2717} & 0\% & 15.3M\\
%     B & \usym{2713} & \usym{2713} & \usym{2717} & -3.56\% & 23.21M\\
%     C & \usym{2713} & \usym{2713} & \usym{2713} & -6.92\% & 24.85M\\
%     \hline
    
%     \end{tabular}
% }
% \vspace{-0.2cm}
% \end{table}
% \vspace{-0.1cm}

\section{Conclusion}
\label{sec:conclusion}

In this paper, we propose a Spatial Importance Guided Variable-rate Image Compression method, called SigVIC. Specifically, a spatial gating unit (SGU) and a spatial scaling network (SSN) are designed for using spatial importance to guide the feature scaling and bit allocation for variable-rate. Besides, a shallow feature fusion module (SFFM) is adopted to refine the decoded features with Top-K shallow features. Experimental results show that our method can yield a state-of-the-art performance, with significantly storage saving and flexibility improvement of compression methods.

\section{Acknowledgement}
\label{sec:ack}

This work was supported in part by the National Natural Science Foundation of China under Grant 61972028, Grant 61902022, and Grant 62120106009.

\vfill\pagebreak

% \section{REFERENCES}
% \label{sec:refs}

% List and number all bibliographical references at the end of the
% paper. The references can be numbered in alphabetic order or in
% order of appearance in the document. When referring to them in
% the text, type the corresponding reference number in square
% brackets as shown at the end of this sentence \cite{C2}. An
% additional final page (the fifth page, in most cases) is
% allowed, but must contain only references to the prior
% literature.

% References should be produced using the bibtex program from suitable
% BiBTeX files (here: strings, refs, manuals). The IEEEbib.bst bibliography
% style file from IEEE produces unsorted bibliography list.
% -------------------------------------------------------------------------
\bibliographystyle{IEEEbib}
\bibliography{strings, refs}

\end{document}